\documentclass[pra,aps,nopacs,nokeys,superscriptaddress,twocolumn,twoside]{revtex4}

\usepackage{graphicx,epic,eepic,epsfig,amsmath,latexsym,amssymb,verbatim,color}

\usepackage{theorem}
\newtheorem{definition}{Definition}
\newtheorem{proposition}[definition]{Proposition}
\newtheorem{lemma}[definition]{Lemma}

\newtheorem{theorem}[definition]{Theorem}

\newtheorem{expl}[definition]{Example}

\def\squareforqed{\hbox{\rlap{$\sqcap$}$\sqcup$}}
\def\qed{\ifmmode\squareforqed\else{\unskip\nobreak\hfil
\penalty50\hskip1em\null\nobreak\hfil\squareforqed
\parfillskip=0pt\finalhyphendemerits=0\endgraf}\fi}
\def\endenv{\ifmmode\;\else{\unskip\nobreak\hfil
\penalty50\hskip1em\null\nobreak\hfil\;
\parfillskip=0pt\finalhyphendemerits=0\endgraf}\fi}

\newlength{\blank}
\newlength{\equalsign}
\newenvironment{beweis}[1][{\hspace{-\blank}}]{{\noindent\emph{Proof~{#1}.\ }}}{\hfill\qed\vskip 0.5\baselineskip}

\mathchardef\ordinarycolon\mathcode`\:
\mathcode`\:=\string"8000
\def\vcentcolon{\mathrel{\mathop\ordinarycolon}}
\begingroup \catcode`\:=\active
  \lowercase{\endgroup
  \let :\vcentcolon
  }

\newcommand{\nc}{\newcommand}
\nc{\rnc}{\renewcommand}
\nc{\beq}{\begin{equation}}
\nc{\eeq}{{\end{equation}}}
\nc{\beqa}{\begin{eqnarray}}
\nc{\eeqa}{\end{eqnarray}}
\nc{\lbar}[1]{\overline{#1}}
\nc{\bra}[1]{\langle#1|}
\nc{\ket}[1]{|#1\rangle}
\nc{\ketbra}[2]{|#1\rangle\!\langle#2|}
\nc{\braket}[2]{\langle#1|#2\rangle}
\nc{\proj}[1]{| #1\rangle\!\langle #1 |}
\nc{\avg}[1]{\langle#1\rangle}
\nc{\Rank}{\operatorname{Rank}}
\nc{\smfrac}[2]{\mbox{$\frac{#1}{#2}$}}
\nc{\tr}{\operatorname{Tr}}
\nc{\ox}{\otimes}
\nc{\dg}{\dagger}
\nc{\dn}{\downarrow}
\nc{\cA}{{\cal A}}
\nc{\cB}{{\cal B}}
\nc{\cC}{{\cal C}}
\nc{\cD}{{\cal D}}
\nc{\cE}{{\cal E}}
\nc{\cF}{{\cal F}}
\nc{\cG}{{\cal G}}
\nc{\cH}{{\cal H}}
\nc{\cI}{{\cal I}}
\nc{\cJ}{{\cal J}}
\nc{\cK}{{\cal K}}
\nc{\cL}{{\cal L}}
\nc{\cM}{{\cal M}}
\nc{\cO}{{\cal O}}
\nc{\cP}{{\cal P}}
\nc{\cR}{{\cal R}}
\nc{\cS}{{\cal S}}
\nc{\cT}{{\cal T}}
\nc{\cX}{{\cal X}}
\nc{\cZ}{{\cal Z}}
\nc{\tA}{{\widetilde{A}}}
\nc{\tB}{{\widetilde{B}}}
\nc{\tE}{{\widetilde{E}}}
\nc{\csupp}{{\operatorname{csupp}}}
\nc{\qsupp}{{\operatorname{qsupp}}}
\nc{\supp}{{\operatorname{supp}\,}}
\nc{\var}{\operatorname{var}}
\nc{\rar}{\rightarrow}
\nc{\lrar}{\longrightarrow}
\nc{\polylog}{\operatorname{polylog}}
\nc{\1}{{\openone}}
\nc{\GHZ}{{\Gamma}}
\nc{\EPR}{{\Phi_2}}

\def\d{\delta}

\nc{\RR}{{{\mathbb R}}}
\nc{\CC}{{{\mathbb C}}}
\nc{\FF}{{{\mathbb F}}}
\nc{\NN}{{{\mathbb N}}}
\nc{\ZZ}{{{\mathbb Z}}}
\nc{\PP}{{{\mathbb P}}}
\nc{\QQ}{{{\mathbb Q}}}
\nc{\UU}{{{\mathbb U}}}
\nc{\EE}{{{\mathbb E}}}
\nc{\id}{{\operatorname{id}}}
\nc{\Span}{{\operatorname{span}}}

\nc{\trho}{{\widetilde{\rho}}}
\nc{\hatphi}{{\widehat{\phi}}}
\nc{\tCN}{{\widetilde{\cal N}}}
\nc{\PEq}{{P^q_{\text{err}}}}
\nc{\PEc}{{P^c_{\text{err}}}}
\nc{\rank}{{\operatorname{rank}}}

\nc{\be}{\begin{equation}}
\nc{\ee}{{\end{equation}}}
\nc{\bea}{\begin{eqnarray}}
\nc{\eea}{\end{eqnarray}}
\nc{\<}{\langle}
\rnc{\>}{\rangle}
\nc{\Hom}[2]{\mbox{Hom}(\CC^{#1},\CC^{#2})}
\nc{\rU}{\mbox{U}}

\nc{\ob}[1]{#1}

\begin{document}

\title{Random quantum codes from Gaussian ensembles and an uncertainty relation}

\author{Patrick Hayden}
\affiliation{School of Computer Science, McGill University, Montreal, Canada}
\email{patrick@cs.mcgill.ca}

\author{Peter W. Shor}
\affiliation{Department of Mathematics, Massachusetts Institute of Technology,
             77 Massachusetts Avenue, Cambridge, MA 02139, USA}
\email{shor@math.mit.edu}

\author{Andreas Winter}
\affiliation{Department of Mathematics, University of Bristol,
             University Walk, Bristol BS8 1TW, U.K.}
\affiliation{Centre for Quantum Technologies, National University of Singapore,
             2 Science Drive 3, Singapore 117542}
\email{a.j.winter@bris.ac.uk}

\date{14 November 2007}

%\date{20th December 2005}

\begin{abstract}
Using random Gaussian vectors and an information-uncertainty relation, 
we give a proof that the coherent information is an achievable rate for
entanglement transmission through a noisy quantum channel. The codes are random subspaces
selected according to the Haar measure, but distorted as a function of the sender's input
density operator. Using large deviations techniques, we show that classical data 
transmitted in either of two Fourier-conjugate bases for the coding subspace can 
be decoded with low probability of error. A recently discovered 
information-uncertainty relation then implies that the quantum mutual information 
for entanglement encoded into the subspace and transmitted through the 
channel will be high. The monogamy of quantum correlations finally implies 
that the environment of the channel cannot be significantly coupled to the 
entanglement, and concluding, which ensures the existence of a decoding by the receiver.
\end{abstract}

\maketitle

\section{Problem and background}
For a bipartite quantum state $\rho^{AB}$, the \emph{coherent information}
is defined to be
\[
  I(A\rangle B)_\rho = H(\rho^B) - H(\rho^{AB}),
\]
where $H$ denotes the von Neumann entropy.
Sometimes, if the state is clear from context, we omit the subscript
and simply write $H(A)$, $I(A\rangle B)$, etc. By way of notation,
we adopt the habit of writing the (Hilbert space) dimension of $A$
as $|A|$.

The \emph{hashing inequality}~\cite{BDSW} is the
statement that asymptotically many copies of $\rho$ have a yield of
$I(A\rangle B)$ ebits per copy under entanglement distillation procedures
with only local operations and one-way classical communication from Alice to Bob. 

Closely related, for a quantum channel (i.e.~a completely positive, trace
preserving -- cptp -- map on density operators)
\[
  {\cal N}:{\cal B}(A') \longrightarrow {\cal B}(B)
\]
and a reference state $\rho^{A'}$ on $A'$, we can define the coherent
information $I_c(\rho;{\cal N})$ of the channel with respect to $\rho$
as follows: Consider a purification $\ket{\phi}^{AA'}$ of $\rho^{A'}$,
and letting $\omega^{AB} := (\id\otimes{\cal N})\proj{\phi}$, define
\[
  I_c(\rho;{\cal N}) = I(A\rangle B)_\omega.
\]
Introducing an isometric Stinespring dilation
\[
  V : A' \hookrightarrow B \otimes E,
\]
for ${\cal N}$
mapping the input Hilbert space $A$ into the combined output and environment
spaces, we can re-express this quantity as follows: introduce the three-party
state
\[
  \ket{\psi}^{ABE} = (\1\otimes V)\ket{\phi}^{AA'},
\]
which is a purification of $\omega^{AB}$. Then
\[
  I_c(\rho;{\cal N}) = H(B)_\psi - H(E)_\psi.
\]

Finally, we need the concept of quantum code: for a
channel $\tCN:{\cal B}(\tA') \rightarrow {\cal B}(\tB)$, this is
given by a pair of cptp encoding and decoding maps
\begin{align*}
  {\cal E}: &\,{\cal B}(\CC^N) \rightarrow {\cal B}(\tA'), \\
  {\cal D}: &\,{\cal B}(\tB) \rightarrow {\cal B}(\CC^N).
\end{align*}
The important parameters of a code are the dimension $N$ of the encoded
system, and the error, given by the trace distance
\[
  \PEq := \left\| ({\cal D} \circ \tCN \circ {\cal E} \otimes \id)\Phi_N - \Phi_N \right\|_1,
\]
where $\Phi_N = \frac{1}{N}\sum_{j,k} \ket{jj}\!\bra{kk}$
is the maximally entangled state on $\CC^N \otimes \CC^N$.
For more on the history of these concepts, motivation, etc., we refer
the reader to the companion papers~\cite{average-norm} and~\cite{privacy-coding};
see also~\cite{tema-con-variazioni}.

The main results we are going to prove are the following two:
\begin{theorem}
  \label{thm:main-oneshot}
  Let $\tCN:{\cal B}(\tA') \rightarrow {\cal B}(\tB)$ be a quantum channel with
  Stinespring dilation $V:\tA' \hookrightarrow \tB\tE$, $\trho$ an input
  density operator, and $P^B$, $P^E$ projections in $\tB$, $\tE$, respectively,
  with the following properties (for some $1/3 \geq \epsilon>0$ and $D,\Delta>0$):
  \begin{align*}
    \tr \bigl( (V\trho V^\dagger) (P^B\otimes P^E) \bigr) &\geq 1-\epsilon, \\
    P^B \tCN(\trho) P^B                                   &\leq D^{-1} P^B, \\
    \trho                                                 &\leq \Delta^{-1}\1.
  \end{align*}
  Then, for $0 < \eta < 1$, there exists a quantum code with encoded dimension
  \[
    N \leq \min \left\{ \eta \frac{D}{\rank P^E}, \eta \Delta \right\},
  \]
  and error $\PEq \leq 2\sqrt{ 2H_2(2\lambda)+4\lambda\log N }$,
  where $H_2(x) = -x\log x - (1-x)\log(1-x)$ is the binary entropy,
  and
  \[
    \lambda = 9\sqrt{\epsilon} + 7\sqrt{\eta} + 3 N \exp(-N\epsilon^2/4).
  \]
  Assuming $N\geq 2$, one obtains the simplified error bound
  \[
    \PEq \leq 7\sqrt{\log N}\sqrt[4]{\lambda}.
  \]
\end{theorem}

A particular case is that of a memoryless channel $\tCN = {\cal N}^{\ox n}$.
We call $Q$ an achievable quantum rate for ${\cal N}$ if there exists a sequence
of codes $({\cal E}_n,{\cal D}_n)$ with input dimensions $N_n$ and error
$\PEq \rightarrow 0$ as $n\rightarrow\infty$, such that
\[
  \liminf_{n\rightarrow\infty} \frac{1}{n} \log N_n \geq Q.
\]
\begin{theorem}[Lloyd~\cite{lloyd:Q}, Shor~\cite{shor:Q} and Devetak~\!\cite{devetak:Q}]
  \label{thm:main-iid}
  Consider a quantum channel ${\cal N}:{\cal B}(A)\rightarrow{\cal B}(B)$,
  and an input state $\rho$ on $A'$. Then, the coherent information 
  $I_c(\rho,{\cal N})$ is an achievable quantum rate.
\end{theorem}

In fact, using the concept of typical subspace, the second 
theorem follows easily from the first. We will prove
Theorem~\ref{thm:main-oneshot} in section~\ref{sec:analysis},
after introducing Gaussian random vectors in section~\ref{sec:gaussian},
and describing the random codes we are going to look at
in section~\ref{sec:random-proj}. The great conceptual significance of
Theorem~\ref{thm:main-iid} is that it makes it possible to express the quantum
capacity of ${\cal N}$, i.e.~the largest achievable rate, in terms of
the coherent information; thanks to a matching upper bound by
Schumacher and Nielsen~\cite{schu:niel}, the capacity is thus given by
\[
  Q({\cal N}) = \lim_{n\rightarrow\infty} \frac{1}{n}
                                          \max_{\rho^{(n)}} I_c(\rho^{(n)};{\cal N}^{\ox n}).
\]

Deducing Theorem~\ref{thm:main-iid} from
Theorem~\ref{thm:main-oneshot} is a straightforward application of
typical subspace techniques~\cite{quantum:coding} -- see appendix A:
choose projectors $P_\delta^A$, $P_\delta^B$, $P_\delta^E$ in $A^n$, $B^n$, $E^n$,
respectively, according to Lemma~\ref{lemma:typical} (appendix A). 
Furthermore, let $\tA = A_\delta$ be the support of $P_\delta^A$,
$\tB=B^n$, $\tE=E^n$ and 
$\trho = \frac{1}{\tr\rho^{\ox n}P_\delta^A} P_\delta^A\rho^{\ox n}P_\delta^A$.
Then the conditions of Theorem~\ref{thm:main-oneshot} are
satisfied, with $\rank P^E = 2^{nH(E)+n\delta}$,
$D = 2^{nH(B)-n\delta}$ and $\Delta = 2^{nH(A)-n\delta}$,
for $\epsilon = 2\cdot 2^{-c n \delta^2}$ and all sufficiently large $n$.
Letting $\gamma = 2^{-c n \delta^2}$, we see that we may
take $N = 2^{nI(A\rangle B)-3n\delta}$, and the get a code
of encoded dimension $N$ and with error exponentially small
in $n$. In other words, the rate $I(A\rangle B)-3\delta$ is
achievable; since $\delta>0$ is arbitrary, Theorem~\ref{thm:main-iid}
follows.
\qed

\medskip
The strategy we will use to prove Theorem~\ref{thm:main-oneshot} will be familiar
from various Shannon-style proofs; we shall find a subspace
of the input space by an appropriate random selection,
However, the analysis of the code differs from the approaches of
the companion papers~\cite{average-norm} and \cite{privacy-coding}.

Both these and the present proof hinge on
the demonstration that the input and environment
of the channel decouple when used with the appropriate code.
Once this decoupling is established, the existence of a decoding/error
correction procedure for the receiver follows by a standard
argument.

So, all three proofs proceed via decoupling of the channel environment
or, equivalently, by forcing the quantum mutual information between
input and environment to be (close to) zero.
This is shown by direct calculation in~\cite{average-norm}.
In~\cite{privacy-coding}, following~\cite{devetak:Q}, one first
shows that the code subspace has a basis such that the receiver
can successfully measure-decode the basis state while the environment
learns (almost) nothing about it -- after which one ``makes the
decoding coherent''.
Here, it is done by not involving the environment at all: instead,
we show that both a special orthonormal basis of the subspace as well as the Fourier
conjugate basis can be decoded at the output. This means that the Holevo quantities
of the two state ensembles, basis and Fourier-conjugate, are
close to maximal, implying, via a recent information-uncertainty
relation, that the quantum mutual information down the channel
is close to maximal. This finally yields the conclusion that the crucial
mutual information between the input and the environment is
close to zero.

We think that this analysis is closest (among the three proofs collected
in this issue) to the original idea in~\cite{shor:Q}. It is still not
the same, as there an explicit description of a quantum decoder is
given, without recourse to decoupling the input from the environment.
See however the recent paper~\cite{klesse:Q} for an alternative argument.

\medskip
The rest of the paper is organised as follows: in section~\ref{sec:gaussian}
we introduce the notion of Gaussian distributed random vectors
(``Gaussian vectors'' for short) and review some of their properties,
mostly cited from~\cite{rsp}, except for a tail bound on the quantum
expectation of random states with an arbitrary observable.
Then, in section~\ref{sec:random-proj}, we define the quantum codes
which we show to be good
quantum transmission codes achieving the bound of Theorem~\ref{thm:main-oneshot}
in section~\ref{sec:analysis}.
Two appendices serve to collect various auxiliary results about
states, measurements, and typical subspaces used throughout the paper,
in addition to miscellaneous proofs.

\section{Gaussian vectors}
\label{sec:gaussian}
We take the following definitions in abridged form from appendix A of~\cite{rsp};
the interested reader is encouraged to consult the referenced paper.

A \emph{Gaussian complex number} with mean $0$ and variance $\sigma^2>0$
is a random variable $X+iY$, where $X$ and $Y$ are independent
real random variables with
$X\sim N\!\left(0,\frac{\sigma^2}{2}\right)$ and
$Y\sim N\!\left(0,\frac{\sigma^2}{2}\right)$.
Its distribution is denoted $N_\CC(0,\sigma^2)$. 

For any orthonormal basis $\{ \ket{1},\ldots,\ket{D} \}$ of $\CC^D$,
a \emph{Gaussian vector} is defined to be a random variable
$\ket{g} \in \CC^D$ whose distribution is described as follows:
\[
  \ket{g} = \sum_{i=1}^D c_i \ket{i},
\]
with $N$ independent Gaussian complex numbers
$c_1,\ldots,c_D \sim N_\CC(0,1/D)$. It is a fundamental property of the
above sum that the resulting distribution is independent of the basis
chosen. I.e., the distribution is unitarily invariant, and in particular,
its density depends only on the length $\| \ket{g} \|_2 = \sqrt{\bra{g} g\rangle}
= \sqrt{\sum_i |c_i|^2}$.
Indeed, we defined the Gaussian vectors in just such a way that
$\EE \bra{g} g\rangle = 1$. And according to Lemma~\ref{lemma:Tr-A}
below the distribution is strongly concentrated around this value.
\begin{lemma}
  \label{lemma:Tr-A}
  Let $\ket{g}$ and $\ket{g_1},\ldots,\ket{g_K}$ be
  independent Gaussian vectors in $\CC^D$.
  Then, for $0\leq \epsilon \leq 1$,
  \[
    \Pr\left\{ |\tr\proj{g} - 1| > \epsilon \right\} \leq 2\exp\bigl( -\epsilon^2 d/6 \bigr),
  \]
  and, for a projector $P$ of rank $r$,
  \begin{equation*}
    \Pr\left\{ \left| \sum_{k=1}^K \tr \proj{g_k}P - \frac{rK}{D} \right| 
                                              > \epsilon \frac{rK}{D} \right\}
                                           \leq 2\exp\left( -rK\frac{\epsilon^2}{6} \right)\!.
  \end{equation*}
  Furthermore, for $\epsilon \leq 1/3$, and $0 \leq A \leq \1$ an operator,
  \begin{align}
    \label{eq:Tr-A-plus}
    \Pr\left\{ \tr \proj{g} A \!>\! (1+\epsilon)\frac{\tr A}{d} \right\}
                                      &\leq \exp\left(\! -\frac{\epsilon^2}{4}\tr A \!\right)\!, \\
    \label{eq:Tr-A-minus}
    \Pr\left\{ \tr \proj{g} A \!<\! (1-\epsilon)\frac{\tr A}{d} \right\}
                                      &\leq \exp\left(\! -\frac{\epsilon^2}{4}\tr A \!\right)\!.
  \end{align}
\end{lemma}
\begin{beweis}
  The first and second statement, about the lengths of Gaussian vectors and average
  inner products, is from Lemma 3 in~\cite{rsp} -- see also
  appendix A there -- or Lemma II.3 in~\cite{rand}.
  
  The third is a generalisation of Lemma 3 in~\cite{rsp}
  (Lemma II.3 in~\cite{rand}). It is proved in appendix B.
\end{beweis}

\section{Random subspace projectors}
\label{sec:random-proj}
For an input space $\tA$ of dimension $|\tA|$, and reference state
$\trho$, the code will be chosen 
as follows: pick a subspace $S_0$ of dimension $N$ according to the
Haar measure, denoting its corresponding subspace projector $P_0$.
Then, let $S = \sqrt{\trho} S_0$, so its subspace projection
$P$ projects onto $\supp \sqrt{\trho} P_0 \sqrt{\trho}$, the support of
the projector $\sqrt{\trho} P_0 \sqrt{\trho}$; this will be our
random code for Theorem~\ref{thm:main-oneshot}.

Our preferred way of describing this random selection is via
a spanning set of vectors drawn independently as follows. 
For $j=1,\ldots,N$, let $\ket{g_j}$ be i.i.d.~Gaussian vectors in $\tA$.
With probability one, these are linearly independent, so they span
an $N$-dimensional subspace $S_0$, which, by the unitary invariance
of the Gaussian measure, is itself distributed according to the unitarily invariant
measure.
Now let
\[
  \ket{\gamma_j} := \sqrt{|\tA|\trho}\,\ket{g_j}.
\]
These vectors will turn out to be almost normalised, with high
probability. They clearly span $S = \sqrt{\trho}S_0$, but we are
after more; we need an orthogonal basis of $S$. To get this, we
follow the recipe of the ``square root'' or ``pretty good'' measurement:
with the (random) operator
$\Gamma := \sum_{j=1}^N \proj{\gamma_j}$, we finally define
\[
  \ket{\phi_j} := \Gamma^{-1/2} \ket{\gamma_j},
\]
which is an orthogonal basis of $S$ (if the $\ket{\gamma_j}$ are
linearly independent) because the subspace projector is
$P = \sum_j \proj{\phi_j}$.

As outlined in the introduction, we will aim to show that this
basis, sent through the channel with equal probabilities,
will yield an output ensemble of states $\sigma_j = {\cal N}(\phi_j)$
with Holevo information close to $\log N$. In fact, we have to show this for the basis
$\{\ket{\phi_j}\}$ as well as for its Fourier-conjugate basis consisting
of the vectors
\[
  \ket{\hatphi_k} = \frac{1}{\sqrt{N}} \sum_j e^{2\pi i jk/N}\ket{\phi_j}.
\]

On the face of it, this set of vectors could have a peculiar, perhaps
hard to describe, distribution. This is not at all the case thanks to
the particular properties of the Gaussian distribution and the
Fourier transform.
\begin{definition}
  \label{defi:conjugate}
  We call a family $\bigl\{ \ket{w_1},\ldots,\ket{w_N} \bigr\}$ of vectors
  \emph{formally Fourier-conjugate} to the family of vectors
  $\bigl\{ \ket{v_1},\ldots,\ket{v_N} \bigr\}$, if for all $k$,
  \[
    \ket{w_k} = \frac{1}{\sqrt{N}} \sum_j e^{2\pi i jk/N} \ket{v_j}.
  \]
  Note that we do not demand normalisation or orthogonality of the vectors
  in either family. Also, the dimension $D$ of the space may be different from $N$.
\end{definition}
\begin{lemma}
  \label{lemma:conjugate}
  If the family $\bigl\{ \ket{w_1},\ldots,\ket{w_N} \bigr\}$ of vectors
  is the formal Fourier-conjugate of the family
  $\bigl\{ \ket{v_1},\ldots,\ket{v_N} \bigr\}$, then for all $j$,
  \[
    \ket{v_j} = \frac{1}{\sqrt{N}} \sum_k e^{-2\pi i jk/N} \ket{w_k}.
  \]
  Furthermore,
  \[
    \sum_j \proj{v_j} = \sum_k \proj{w_k}.
  \]
  Finally, if $\bigl\{ \ket{v_1},\ldots,\ket{v_N} \bigr\}$ are independent
  Gaussian vectors with $N\leq D$, then so are 
  $\bigl\{ \ket{w_1},\ldots,\ket{w_N} \bigr\}$.
\end{lemma}
\begin{beweis}
  Straightforward calculations.
\end{beweis}

This means that there is another, equivalent, way of arriving at the
basis $\{\ket{\hatphi_k}\}$ of $S$: namely, start with the set of
(by Lemma~\ref{lemma:conjugate}, Gaussian!) vectors
\[
  \ket{\widehat{g}_k} = \frac{1}{\sqrt{N}} \sum_j e^{2\pi i jk/N} \ket{g_j},
\]
formally Fourier-conjugate to the $\ket{g_j}$. Then we can form the
vectors $\ket{\widehat{\gamma}_k} = \sqrt{|\tA|\trho}\ket{\widehat{g}_k}$,
and they are clearly formally Fourier-conjugate to the $\ket{\gamma_j}$.
Finally, by Lemma~\ref{lemma:conjugate} above, the normalisation
operator $\widehat{\Gamma} = \sum_k \proj{\widehat{\gamma}_k}$
equals $\Gamma$, so we find that
\[
  \ket{\hatphi_k} = \widehat{\Gamma}^{-1/2} \ket{\widehat{\gamma}_k}
                  = \Gamma^{-1/2} \ket{\widehat{\gamma}_k}.
\]

In other words, we have arrive at the
\begin{proposition}
  \label{prop:phi-hatphi}
  The distribution of the set $\{ \ket{\hatphi_k} \}_k$ is
  exactly the same as that of the set $\{ \ket{\phi_j} \}_j$.
  \qed
\end{proposition}

\section{Performance analysis}
\label{sec:analysis}
In the previous section we have described a random subspace
$S$ of $\tA'$. The encoder of the code will simply be the
isometric identification of $\CC^N$ with $S$: ${\cal E} = U\cdot U^\dagger$,
with
\begin{align*}
  U: \CC^N   &\longrightarrow S \hookrightarrow \tA', \\
     \ket{j} &\longmapsto \ket{\phi_j}.
\end{align*}

Following Devetak~\cite{devetak:Q} -- see Lemma 1.1 in~\cite{average-norm} --
we do not worry about the decoding map; it will exist once the
``decoupling from the environment'' condition holds. Namely, denoting
$R=\CC^N$, $\tau^R$ the maximally mixed state on $R$, and
\[
  \ket{\Psi}^{R\tB\tE} := (\1 \ox VU)\ket{\Phi_N},
\]
we know that a decoder ${\cal D}$
with error $p$ exists once we ascertain that
\[
  \left\| \Psi^{R\tE} - \tau^R \ox \vartheta^\tE \right\|_1 \leq p,
\]
for an arbitrary state $\vartheta^\tE$ of the environment.

By Pinsker's inequality~\cite{OhyaPetz} for the relative entropy,
applied to $\Psi^{R\tE}$ and $\tau^R \ox \Psi^\tE$,
\[\begin{split}
  I(R:\tE) &=    D\left( \Psi^{R\tE} \| \tau^R \ox \Psi^\tE \right)  \\
           &\geq \left( \frac{1}{2} \left\| \Psi^{R\tE}
                                           - \tau^R \ox \vartheta^\tE \right\|_1 \right)^2,
\end{split}\]
so it is enough to show $I(R:\tE) \leq p^2/4$.
Here, $I(R:\tE) = H(R)+H(\tE)-H(R\tE)$ is the quantum mutual information,
and $D(\rho\|\sigma) = \tr\rho(\log\rho-\log\sigma)$ is the quantum
relative entropy.

By the elementary identity
\[
  2H(R) = I(R:\tE) + I(R:\tB),
\]
which holds for any pure state on $R\tilde{B}\tilde{E}$,
and with $H(R) = \log N$ in our case, we will be done as soon as
we show $I(R:\tB) \geq 2\log N - p^2/4$. The proof that this inequality holds
for a random subspace is based on 
the following ``information-uncertainty
relation'':
\begin{lemma}[Information-uncertainty~\cite{CW05}, Lemma 1]
  \label{lemma:uncert}
  Let $\cE_0 = \{1/N, \proj{j} \}$ be the uniform
  ensemble for an arbitrary fixed orthonormal basis $\{ \ket{j} \}$ of an
  $N$-dimensional Hilbert space $S$,
  and $\cE_1 = \{1/N, {\rm QFT}\proj{j}{\rm QFT}^\dagger \}$,
  where ${\rm QFT}$ is the Fourier transform in dimension $N$.

  Then, for any quantum channel ${\cal M}$ with input space $S$
  and output $B$,
  \[
    \chi\bigl({\cal M}(\cE_0)\bigr) + \chi\bigl({\cal M}(\cE_1)\bigr)
                                                                     \leq I(R:B)_\omega.
  \]
  Here, the right hand side is the quantum mutual information of the
  state $\omega^{RB} = (\id\ox{\cal M})\Phi_d$, where $\Phi_d$ is the
  maximally entangled state on $RS$.
  On the left hand side, we have two Holevo informations~\cite{Holevo}
  of the ensembles ${\cal M}({\cal E}_i)$ of channel output states;
  for an arbitrary ensemble ${\cal E} = \{ p_x,\sigma_x \}$ of
  states,
  \[
    \chi({\cal E}) := H\left( \sum_x p_x\sigma_x \right) - \sum_x p_x H(\sigma_x).
  \]
  \qed
\end{lemma}

Of course, the assumption of this lemma is just our situation: we have
a subspace $S$ of dimension $N$ in $\tA$, and consider two Fourier-conjugate
bases.

Hence, in the light of Proposition~\ref{prop:phi-hatphi}, all we need
to show is the following:
\begin{proposition}
  \label{prop:main}
  Under the assumptions of Theorem~\ref{thm:main-oneshot}, 
  consider independent Gaussian vectors $\ket{g_1},\ldots,\ket{g_N} \in \tA'$,
  as well as
  \begin{align*}
    \ket{\gamma_j} &:= \sqrt{|\tA|\trho}\,\ket{g_j}, \\
    \Gamma         &:= \sum_j \proj{\gamma_j},     \\
    \ket{\phi_j}   &:= \Gamma^{-1/2} \ket{\gamma_j}.
  \end{align*}
  Then, for the output ensemble
  \[
    {\cal E} = \left\{ 1/N, \sigma_j := \tCN(\proj{\phi_j}) \right\},
  \]
  it holds with probability $>1/2$ that
  \[
    \chi({\cal E}) \geq \log N - H_2(2\lambda) - 2\lambda\log N,
  \]
  where
  \[
    \lambda = 9\sqrt{\epsilon} + 7\sqrt{\eta} + 3N \exp( - N\epsilon^2/6 ).
  \]
\end{proposition}

As a consequence, we have that with positive probability both ${\cal E}$
and the ensemble obtained from the Fourier-conjugate inputs,
\[
  \widehat{\cal E} = \left\{ 1/N, \tCN(\proj{\hatphi_k}) \right\},
\]
have $\chi({\cal E}),\ \chi(\widehat{\cal E}) \geq \log N - H_2(2\lambda) - 2\lambda\log N$.
By Lemma~\ref{lemma:uncert} this means
$I(R:\tB) \geq 2\log N - 2H_2(2\lambda) - 4\lambda\log N$,
hence $I(R:\tE) \leq 2H_2(2\lambda) + 4\lambda\log N$, and we are done.
Observing that $H_2(x) \leq 2\sqrt{x(1-x)}$, the right hand side
can be further upper bounded by $6\sqrt{\lambda} + 4\lambda\log N$,
which is $\leq 10\sqrt{\lambda}\log N$ as long as $N \geq 2$.

To conclude, we use Pinsker's inequality, as described at the start
of this section, to relate $\PEq$ and (the upper bounds on) the
mutual information $I(R:\tE)$.

\medskip
\begin{beweis}[of Proposition~\ref{prop:main}]
  What we shall show is that there exists a classical decoder for the ensemble
  achieving small error probability; i.e. we need to find a POVM
  $(\Lambda_j)_{j=1}^N$ such that
  \[
    \PEc : = \frac{1}{N} \sum_j \tr\bigl[ \sigma_j (\1-\Lambda_j) \bigr]
  \]
  is small, at least in expectation. Then, denoting the random output of the
  measurement $j'$, we have that by the monotonicity of
  the Holevo quantity under post-processing and the classic Fano
  inequality~\cite{cover:thomas},
  \[
    \chi({\cal E}) \geq I(j:j') \geq \log N - \PEc \log N - H_2(\PEc).
  \]
  Looking at this, we are done once we show that
  \[
    \EE \PEc \leq 9\sqrt{\epsilon} + 7\sqrt{\eta} + 3N \exp( - N\epsilon^2/6 ) =: \lambda.
  \]
  The reason is Markov's inequality, telling us that the probability
  of a random random code having $\PEc > 2\lambda$ is strictly
  smaller than $1/2$.

  For this, we first analyse random codes drawn from the ensemble
  $\ket{\gamma} = \sqrt{|\tA|\trho}\,\ket{g}$, with Gaussian $\ket{g}$.
  The states $\gamma = \proj{\gamma}$ and so the $\sigma_g := \tCN(\gamma)$ are of
  course not generally normalised, but we can still apply the
  Packing Lemma (Lemma~\ref{lemma:packing}
  in appendix A). There, we let $\Pi = P^B$, and the $\Pi_g$
  for the individual ensemble states $\tCN(\gamma)$ are constructed
  as follows: observe that 
  $\ket{\gamma'} := (\1 \ox P^E)V\ket{\gamma} \in \tB \ox \tE$
  is a vector of Schmidt rank at most $d = \rank P^E$, so we may
  choose $\Pi_g$ to be the projector onto the support of
  $\tr_E \proj{\gamma'}$. The conditions of the Packing Lemma
  are easily verified -- observing that $\EE \gamma = \trho$,
  so $\EE \sigma_g = \tCN(\trho) =: \sigma$.
  
  We conclude that for i.i.d.~$\{ \ket{\gamma_1},\ldots,\ket{\gamma_N} \}$
  there is a POVM $\{ \Lambda_1,\ldots,\Lambda_N \}$ such that
  \[
    \EE \PEc \left(  \{{\cal N}(\gamma_j),\Lambda_j \}_{j=1}^N \right)
                                            \leq 6\sqrt{\epsilon} + 4\eta.
  \]
  Now, if we use the same decoder instead for the states
  $\ket{\phi_j} = \Gamma^{-1/2}\ket{\gamma_j}$, we incur additional
  errors, as follows:
  
  First of all, by Lemma~\ref{lemma:Tr-A} applied to $A = \Delta\trho$
  we have, except with probability $\leq 2N \exp( -\Delta\epsilon^2/4 )$, that
  \begin{equation}
    \label{eq:length-bounds}
    \forall j\quad 1-\epsilon \leq \bra{\gamma_j}\gamma_j\rangle \leq 1+\epsilon.
  \end{equation}
  which we shall assume to hold from now on.
  
  Furthermore, we have, using the elementary inequality
  $\| \phi-\gamma \|_1 \leq \sqrt{2}\| \phi-\gamma \|_2$
  for rank one projectors $\phi$ and $\gamma$, and eq.~(\ref{eq:length-bounds}), that
  \begin{equation}\begin{split}
     \frac{1}{N} \sum_j &\frac{1}{2}\| \phi_j - \gamma_j \|_1
              \leq \frac{1}{N} \sum_j \frac{\sqrt{2}}{2} \| \phi_j - \gamma_j \|_2       \\
             &=    \frac{1}{N} \sum_j \frac{1}{\sqrt{2}}
                                          \sqrt{ \bra{\phi_j}\phi_j\rangle^2
                                                +\bra{\gamma_j}\gamma_j\rangle^2
                                               -2|\bra{\phi_j}\gamma_j\rangle|^2 }           \\
             &\leq \frac{1}{N} \sum_j \sqrt{ (1+\epsilon)^2
                                            -|\bra{\phi_j}\gamma_j\rangle|^2 }               \\
             &\leq \sqrt{ \frac{1}{N} \sum_j \left( (1+\epsilon)^2
                                                   -|\bra{\phi_j}\gamma_j\rangle|^2 \right) } \\
             &\leq \sqrt{ \frac{1}{N} \sum_j 2(1+\epsilon)
                                             \bigl( (1+\epsilon)
                                                   -|\bra{\phi_j}\gamma_j\rangle| \bigr) }
    \label{eq:upper-1}
  \end{split}\end{equation}
  where the second-to-last line follows by the concavity of the square root function,
  and the last involves the Cauchy-Schwarz inequality.

  We shall concentrate for the moment on the average under the square root:
  \begin{equation}\begin{split}
     \frac{1}{N} \sum_j &\bigl( (1+\epsilon) - |\bra{\phi_j}\gamma_j\rangle| \bigr)
           = \epsilon + \frac{1}{N} \sum_j \bigl( 1 - |\bra{\phi_j}\gamma_j\rangle| \bigr) \\
          &= \epsilon + \frac{1}{N} \sum_j
                                    \bigl( 1 - \bra{\gamma_j}\Gamma^{-1/2}\ket{\gamma_j} \bigr) \\
          &= \epsilon + 1 - \frac{1}{N}\tr\sqrt{\Gamma},
    \label{eq:upper-2}
  \end{split}\end{equation}
  where we have inserted the definition of the $\ket{\phi_j}$, and noted
  that the inner products $\bra{\phi_j}\gamma_j\rangle$ are non-negative.
  Now, we use a trick from~\cite{HJSWW}: for the positive semidefinite
  operator $\Gamma$,
  \[
    \sqrt{\Gamma} \geq \frac{3}{2}\Gamma - \frac{1}{2}\Gamma^2,
  \]
  so we can continue upper bounding as follows, using
  the abbreviation $S_{jk} = \bra{\gamma_j}\gamma_k\rangle$:
  \[\begin{split}
    1 - \frac{1}{N}\sqrt{\Gamma}
           &\leq 1 - \frac{1}{N}\left( \frac{3}{2}\Gamma - \frac{1}{2}\Gamma^2\right) \\
           &=    \frac{1}{N}\left( N - \frac{3}{2}\sum_j S_{jj}
                                     + \frac{1}{2}\sum_{jk} |S_{jk}|^2 \right)        \\
           &=    \frac{1}{N}\sum_j \left( 1 - \frac{3}{2}S_{jj} + \frac{1}{2}S_{jj}^2 \right)
                  +\frac{1}{N}\sum_{j\neq k} |S_{jk}|^2                               \\
           &=    \frac{1}{N}\sum_j ( 1 - S_{jj} ) \left(1 - \frac{1}{2}S_{jj} \right)
                  +\frac{1}{N}\sum_{j\neq k} |S_{jk}|^2.
  \end{split}\]
  Here, the first term is bounded above by
  $\epsilon \frac{1+\epsilon}{2}$.
  The second term consists of an average of $N$ expressions, one for each $j$, of the form
  \[\begin{split}
    \sum_{k \neq j} |\bra{\gamma_j}\gamma_k\rangle|^2
      &=    \sum_{k \neq j} \left| \bra{\gamma_j}\sqrt{|\tA|\trho}\,\ket{g_k} \right|^2 \\
      &\leq (1+\epsilon)\frac{|\tA|}{\Delta} \sum_{k \neq j} \tr \proj{g_k} P_j,
  \end{split}\]
  with a rank one projector $P_j$. So we can apply Lemma~\ref{lemma:Tr-A}
  once more to find that, except with probability
  $\leq N \exp( -N\epsilon^2/6 )$, the latter 
  expressions are all upper bounded by
  \[
    (1+\epsilon)\frac{|\tA|}{\Delta}\, (1+\epsilon)\frac{N}{|\tA|}
                                                     \leq (1+\epsilon)^2 \eta.
  \]
  Inserting all this into eq.~(\ref{eq:upper-2}), we find
  \[
    \frac{1}{N} \sum_j \bigl( (1+\epsilon) - |\bra{\phi_j}\gamma_j\rangle| \bigr)
        \leq \epsilon + \epsilon\frac{1+\epsilon}{2} + (1+\epsilon)^2 \eta.
  \]
  In turn plugging that into eq.~(\ref{eq:upper-1}), we arrive at
  \[\begin{split}
    \frac{1}{N} \sum_j \frac{1}{2}\| \phi_j - \gamma_j \|_1
      &\leq \sqrt{ 2(1+\epsilon)\left( \epsilon\frac{3+\epsilon}{2} + (1+\epsilon)^2 \eta \right) } \\
      &\leq \sqrt{ 9\epsilon + 9\eta}
       \leq 3\sqrt{\epsilon} + 3\sqrt{\eta},
  \end{split}\]
  remembering $\epsilon \leq 1/3$.

  Putting all this together, with the monotonicity of the trace
  norm under cptp maps and using
  $\tr\bigl( (\rho-\sigma) \Lambda) \leq \frac{1}{2}\| \rho-\sigma \|_1$
  for states $\rho$, $\sigma$ and $0 \leq \Lambda \leq \1$, leads to
  \[\begin{split}
    \EE &\PEc \left(  \{{\cal N}(\phi_j),\Lambda_j \}_{j=1}^N \right)                \\
        &\leq \EE \PEc \left(  \{{\cal N}(\gamma_j),\Lambda_j \}_{j=1}^N \right)     \\
        &\phantom{=======}
                       + 3\sqrt{\epsilon} + 3\sqrt{\eta} + 3N \exp( -N\epsilon^2/6 ) \\
        &\leq 6\sqrt{\epsilon} + 4\eta 
                       + 3\sqrt{\epsilon} + 3\sqrt{\eta} + 3N \exp( -N\epsilon^2/6 ),
  \end{split}\]
  and we are done.
\end{beweis}

\section{Conclusion}
\label{sec:conclusion}
We have given yet another proof of the direct part of the quantum channel coding theorem,
in the sense of showing the achievability of the coherent information rate.

The present proof is distinguished from other approaches in that it
is shown that the classical information in two Fourier-conjugate
bases of the code subspace can be recovered at the output. Application
of a recent information-uncertainty relation then ensures that
the quantum information in the subspace can in fact be decoded.

It is tempting to speculate that the role of the pair of measurement-decoders
for the two conjugate bases is to implement the measurement of the familiar basis
and phase errors of a conventional quantum
error correcting code, or their equivalents. To give more substance to this idea,
it would be necessary to show how to build the quantum decoder directly from the
two measurement-decoders. We leave this as an open problem.

\acknowledgments
We would like to thank Micha\l{} Horodecki for many stimulating discussions
on the proof(s) of the quantum channel coding theorem, and his insistence
that the present proof should be written up and published.

PH is supported by the Canada Research
Chairs program, CIFAR, FQRNT, MITACS, NSERC and QuantumWorks.
He is also grateful to the DAMTP in Cambridge for their hospitality. 
PWS is partially supported by the W. M. Keck Foundation Center for
Extreme Quantum Information Theory, and through the National Science
Foundation through grant CCF-0431787.
AW is supported by the U.K. EPSRC (project ``QIP IRC'' and an Advanced
Research Fellowship), by a Royal Society Wolfson Merit Award, and
the EC, IP ``QAP''. The Centre for Quantum Technologies is funded by the
Singapore Ministry of Education and the National Research Foundation as
part of the Research Centres of Excellence programme.

\appendix

\section{Miscellaneous Lemmas}
\label{app:misc}

%\begin{lemma}[Fannes inequality~\cite{fannes}]
%  \label{lemma:fannes}
%  For two states $\rho$, $\sigma$ on a $d$-dimensional space, such
%  that $\| \rho-\sigma \|_1 \leq \epsilon \leq 1/e$,
%  \[
%    | H(\rho)-H(\sigma) | \leq \epsilon \log d - \epsilon\log\epsilon.
%  \]
%  \qed
%\end{lemma}

\begin{lemma}[Packing~\cite{devetak-packing}]
  \label{lemma:packing}
  Consider an  ensemble $\{p_m, \sigma_m \}$ of positive semidefinite
  operators (not necessarily states!) with average 
  $\sigma = \sum_{m} p_m \sigma_m$, which is assumed to be
  a density operator; in particular, $\sum_m p_m \tr\sigma_m = 1$.
  Assume the existence of projectors $\Pi$ and $\Pi_m$
  with the following properties:
  \begin{align*}
    \sum_m p_m \tr \sigma_m \Pi_m & \geq  1 - \epsilon, \\
    \sum_m p_m \tr \sigma_m \Pi   & \geq  1 - \epsilon, \\
    \tr \Pi_m          & \leq  d,            \\
    \Pi \sigma \Pi     & \leq  D^{-1} \Pi,
  \end{align*}
  for all $m$.
  Let $N = \lfloor \eta D/d \rfloor$ for some $0 < \eta < 1$, and
  pick $m_1,\ldots,m_N$ independently at random according to the distribution
  $p_m$.
  
  Then there exists a corresponding
  POVM $\{\Lambda_k\}_{k=1}^N$ which reliably distinguishes between the states
  $\{\sigma_{m_k}\}_{k=1}^N$ in the sense that the expectation of
  the (average) error probability of the code $\{\sigma_{m_k},\Lambda_k\}_{k=1}^N$,
  \[
    \PEc = \PEc(\{\sigma_{m_k},\Lambda_k\}) 
         := \frac{1}{N} \sum_k \tr\bigl[ \sigma_{m_k}(\1-\Lambda_k) \bigr],
  \]
  satisfies
  \[
    \EE \PEc \leq 2\epsilon + 4\sqrt{\epsilon} + 4\eta
             \leq 6\sqrt{\epsilon} + 4\eta.
  \]
  (In particular, there exists a code with error bounded by the
  above quantity.)
  
  The same statements hold for continuous ensembles -- the above
  formulation with a discrete probability distribution was chosen
  only for notational convenience.
  \qed
\end{lemma}
\begin{beweis}
  It is almost the same statement and proof as Lemma 2 in~\cite{devetak-packing},
  which itself is an adaptation of a result by Hayashi and Nagaoka~\cite{hayashi:nagaoka}.
  
  Note that we demand state normalisation of the $\sigma_m$ not individually,
  but only in the ensemble average -- which makes the lemma more suitable
  to be applied with the, generally unnormalised, Gaussian input states.
  Inspecting the proof in~\cite{devetak-packing}, it is evident that
  in fact only that is required.

  There are only the following two other differences. We use the slightly
  better ``Gentle measurement Lemma'' of Ogawa and Nagaoka~\cite{ogawa:nagaoka}
  instead of~\cite{winter:qstrong} -- see Lemma~\ref{lemma:gentle} below.
  And whereas~\cite{devetak-packing} demands that for all $m$,
  \[
    \tr \sigma_m \Pi_m,\ \tr \sigma_m \Pi \geq  1 - \epsilon,
  \]
  our conditions on $\Pi$ and the $\Pi_m$ require this to hold only
  on average over the ensemble $\{p_m, \sigma_m : m \in {\cal M}\}$.
  Looking at the proof in~\cite{devetak-packing}, it is evident
  that this condition is indeed enough for the conclusion.
\end{beweis}

\begin{lemma}[Gentle measurement~\cite{winter:qstrong}
              and~\cite{ogawa:nagaoka}]
  \label{lemma:gentle}
  Let $\rho$ be positive semidefinite, and $0 \leq X \leq \1$ be an operator on
  some Hilbert space, such that $\tr \bigl(\rho (\1-X)\bigr) \leq \epsilon \tr\rho$.
  Then,
  \[
    \bigl\| \rho - \sqrt{X}\rho\sqrt{X} \bigr\|_1 \leq 2\sqrt{\epsilon} \tr\rho.
  \]
  \qed
\end{lemma}

%\begin{lemma}
%  \label{lemma:H2-sqrt}
%  The binary entropy $H_2(x) = H(x,1-x) = -x\log x - (1-x)\log(1-x)$
%  satisfies, for $0\leq x \leq 1$,
%  \[
%    H(x,1-x) \leq 2\sqrt{x(1-x)}.
%  \]
%\end{lemma}
%\begin{beweis}
%  Elementary.
%\end{beweis}
%
%\medskip

Here follow some properties of typical subspaces as defined
in~\cite{quantum:coding}; we quote directly from~\cite{average-norm}.
Consider a density matrix with spectral decomposition
$\rho^{A} = \sum_x p_x\proj{x}^A$.  Its $n$th tensor power can be written as 
\[
  (\rho^{A})^{\ox n} = \sum_{x^n}p_{x^n}\proj{x^n}^{A^n},
\]
where $p_{x^n} = p_{x_1}\cdots p_{x^n}$ and $\ket{x^n}^{A^n} 
               = \ket{x_1}^A\cdots \ket{x_n}^A$.  
The $\d$-(entropy) typical subspace $A_\d < A^n$ is defined as 
\[
  A_\d = \text{span}\left\{ \ket{x^n}^{A^n} :
             \left| -\frac{1}{n} \log p_{x^n} - H(\rho^{A}) \right| \leq \d \right\},
\]  
and the \emph{$\d$-typical projection} $P_\d^A$ is defined to project
$A^n$ onto $A_\d$.  We shall need the following lemma:
\begin{lemma}[Typicality]
  \label{lemma:typical}
  Let a tripartite pure state $\ket{\psi}^{ABC}$ be given.  
  For every $\delta > 0$ and all sufficiently large $n$ there 
  are $\d$-typical projections $P_\d^A$, $P_\d^B$ and $P_\d^E$ onto 
  $\d$-typical subspaces $A_\d\subseteq A^n$, $B_\d\subseteq B^n$ 
  and $E_\d\subseteq E^n$, respectively, such that the states 
  \begin{align*}
    \ket{\psi}^{A^nB^nE^n}    &:= (\ket{\psi}^{ABE})^{\ox n},  \\
    \ket{\psi_\d}^{A^nB^nE^n} &:= (P_\d^A\ox P_\d^B\ox P_\d^E)\ket{\psi}^{A^nB^nE^n}
  \end{align*} 
  satisfy 
  \begin{align*}
    | A_\delta | &\leq 2^{nH(A) + n\delta}, \\
    | B_\delta | &\leq 2^{nH(B) + n\delta}, \\
    | E_\delta | &\leq 2^{nH(E) + n\delta}, \\
    P_\delta^B \psi^{B^n} P_\delta^B &\leq 2^{-nH(B) + n\delta} P_\delta^B, \\
    \| \psi^{A^nB^nE^n} - \psi_\d^{A^nB^nE^n} \|_1 &\leq \epsilon,
  \end{align*}
  where $\epsilon=2^{-c n \d^2}$ for some constant $c> 0$ independent of $\delta$ and $n$.
\end{lemma}
\begin{beweis}
  See \cite{HOW05b}.
\end{beweis}

\section{Proof of Lemma~\ref{lemma:Tr-A},~eqs.~(\ref{eq:Tr-A-plus})~and~(\ref{eq:Tr-A-minus})}
\label{app:proof}
We shall use the following easy lemma:
\begin{lemma}
  \label{lemma:lem}
  Let $\delta < 1$. Then:
  \begin{align*}
    \text{for } -\delta \leq x \leq 0, &\quad
                                        \ln(1+x) \geq x - \frac{x^2}{2}\frac{1}{1-\delta}; \\
    \text{for } 0\leq x \leq 1,        &\quad
                                        \ln(1+x) \geq x - \frac{x^2}{2}.
  \end{align*}
\end{lemma}
\begin{beweis}
  By Taylor expansion,
  $\ln(1+x) = x - \frac{x^2}{2} + \frac{x^3}{3} - \frac{x^4}{4} + \frac{x^5}{5} \mp \ldots$.
  
  The second bound is the easier one: just group each (positive) odd term
  with its immediately consecutive (negative) even term, i.e.
  \[
    \frac{x^3}{3} - \frac{x^4}{4},\ \frac{x^5}{5} - \frac{x^6}{6},\text{ etc.},
  \]
  all of which are clearly non-negative, and we are done.
  
  For the first bound, write $y = -x \leq \delta$, and observe
  \[\begin{split}
    \ln(1+x) &=    \ln(1-y) = -y - \frac{y^2}{2} - \frac{y^3}{3} - \ldots \\
             &=    -y - \frac{y^2}{2}\left( 1 + \frac{2}{3}y + \frac{2}{4}y^2 + \ldots \right) \\
             &\geq -y - \frac{y^2}{2}\left( 1+y+y^2+y^3+\ldots \right)                         \\
             &=     x - \frac{x^2}{2}\frac{1}{1-y}
              \geq  x - \frac{x^2}{2}\frac{1}{1-\delta}.
  \end{split}\]
\end{beweis}

\bigskip
\begin{beweis}[of the probability bounds~(\ref{eq:Tr-A-plus})~and~(\ref{eq:Tr-A-minus})]
Write $A$ in its eigenbasis,
$A = \sum_i a_i \proj{i}$, with $0 \leq a_i \leq 1$. The Gaussian
vector is $\ket{g} = \sum_i c_i \ket{i}$, with $c_i \sim {\cal N}_{\CC}(0,1/D)$.
Then $\tr(\proj{g} A) = \sum_i a_i|c_i|^2$ is a weighted sum of
independent random variables -- which is where the large deviation
behaviour will come from.

The ``Bernstein trick'' is the realisation that (for $t>0$)
\[\begin{split}
  \Pr&\left\{ \sum_i a_i|c_i|^2 > (1+\epsilon) \frac{\tr A}{D} \right\} \\
      &=    \Pr\left\{ e^{t\sum_i a_i|c_i|^2} > e^{t(1+\epsilon)(\tr A)/D} \right\} \\
      &\leq \left( \EE e^{t\sum_i a_i|c_i|^2} \right) e^{-t(1+\epsilon)(\tr A)/D}   \\
      &=    \prod_i \left( \EE e^{t a_i|c_i|^2} \right) e^{-t(1+\epsilon)a_i/D},
\end{split}\]
the second line by Markov's inequality, and the third by independence of the $c_i$.
We take the evaluation of the expectation above
(known as ``moment generating function'') from~\cite{rsp},
Lemma 23 (appendix A): for $t < D/a_i$,
\[
  \EE e^{t a_i|c_i|^2} = \frac{1}{1-t \frac{a_i}{D}}.
\]
Plugging this in and letting $t = D\frac{\epsilon}{1+\epsilon}$, we get the upper
bound on the probability in question, of
\[
  \prod_i e^{-\epsilon a_i - \ln\left( 1-\frac{\epsilon a_i}{1+\epsilon} \right)}.
\]
The exponents can be upper bounded using Lemma~\ref{lemma:lem}: because
we assume $\epsilon\leq 1/3$, the argument $\frac{\epsilon a_i}{1+\epsilon}$
is bounded above by $(1/3)/(1+1/3) = 1/4$, so we get
\[\begin{split}
  -\epsilon a_i - \ln&\left( 1-\frac{\epsilon a_i}{1+\epsilon} \right) \\
      &\leq -\epsilon a_i +\frac{\epsilon a_i}{1+\epsilon} 
                +\frac{1}{2}\frac{1}{1-1/4}\left( \frac{\epsilon a_i}{1+\epsilon} \right)^2 \\
      &=    -\frac{\epsilon^2 a_i}{1+\epsilon}
                +\frac{2}{3}\frac{\epsilon^2 a_i^2}{(a+\epsilon)^2} \\
      &\leq -\frac{1}{3(1+\epsilon)}\epsilon^2 a_i 
       \leq -\frac{1}{4}\epsilon^2 a_i.
\end{split}\]
So, we finally get that the probability in~(\ref{eq:Tr-A-plus}) is upper bounded by
\[
  \prod_i e^{-\frac{1}{4}\epsilon^2 a_i} = e^{-\frac{\epsilon^2}{4}\tr A},
\]
which is what we wanted.

The bound in the other direction is fairly similar:
here we have, for $t>0$, and pretty much as before (noting that
the extra minus sign reverses the direction of the inequality),
\[\begin{split}
  \Pr&\left\{ \sum_i a_i|c_i|^2 < (1-\epsilon) \frac{\tr A}{D} \right\} \\
      &=    \Pr\left\{ e^{-t\sum_i a_i|c_i|^2} > e^{-t(1-\epsilon)(\tr A)/D} \right\} \\
      &\leq \left( \EE e^{-t\sum_i a_i|c_i|^2} \right) e^{t(1-\epsilon)(\tr A)/D}   \\
      &=    \prod_i \left( \EE e^{-t a_i|c_i|^2} \right) e^{t(1-\epsilon)a_i/D}   \\
      &=    \prod_i \frac{1}{1+t \frac{a_i}{D}} e^{t(1-\epsilon)a_i/D} \\
      &=    \prod_i e^{t(1-\epsilon)a_i/D -\ln\left( 1+ t\frac{a_i}{D} \right)}.
\end{split}\]
Now, choosing $t = D\frac{\epsilon}{1-\epsilon}$, the exponent for each $i$ is
\[\begin{split}
  \epsilon a_i - \ln\left( 1+\frac{\epsilon a_i}{1-\epsilon} \right)
      &\leq \epsilon a_i -\frac{\epsilon a_i}{1-\epsilon} 
                +\frac{1}{2}\left( \frac{\epsilon a_i}{1-\epsilon} \right)^2 \\
      &=    -\frac{\epsilon^2 a_i}{1-\epsilon}
                +\frac{1}{2}\frac{\epsilon^2 a_i^2}{(1-\epsilon)^2}          \\
      &\!\!\!\!\!\!\!\!
       \leq -\epsilon^2 a_i \left( 1 - \frac{1}{2(1-1/3)} \right)
       =    -\frac{1}{4}\epsilon^2 a_i,
\end{split}\]
where we have once more invoked Lemma~\ref{lemma:lem} and used $\epsilon \leq 1/3$.
\end{beweis}

\end{document}